\documentclass[epj]{svjour}
\usepackage{graphicx}
\usepackage{graphics}
\usepackage{color}
\newcommand{\dhat}{\hat{\vec d}}

\begin{document}
\title{Wide shear zones and the spot model: Implications from the split-bottom geometry}

\author{Erik Woldhuis\inst{1} \and Brian P. Tighe\inst{1} \and Wim van Saarloos\inst{1}}

\institute{                    
  \inst{1} Instituut-Lorentz, Universiteit Leiden - Postbus 9506, 2300 RA Leiden, The Netherlands
}

\abstract{
The spot model has been developed by Bazant and co-workers to describe quasistatic granular flows. It assumes that granular flow is caused by the opposing flow of so-called spots of excess free volume, with spots moving along the slip lines of Mohr-Coulomb plasticity. The model is two-dimensional and has been successfully applied to a number of different geometries. In this paper we investigate whether the spot model in its simplest form can describe the wide shear zones observed in experiments and simulations of a  Couette cell with split bottom. We give a general argument that is independent of the particular description of the stresses, but which shows that the present formulation of the spot model in which diffusion and drift terms are postulated to balance on length scales of order of the spot diameter, i.e. of order 3-5 grain diameters, is difficult to reconcile with  the observed wide shear zones. We also discuss the implications for the spot model of co-axiality of the stress and strain rate tensors found in these wide shear flows, and point to possible extensions of the model that might allow one to account for the existence of wide shear zones.
} 
\maketitle

\section{Introduction}
\begin{figure}
        \begin{center}
          \includegraphics[scale=0.25]{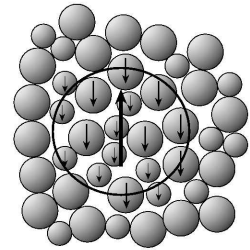}      
        \end{center}
      \caption{Caricature \cite{bazant06a,bazant06b} of a spot, a region where the density of grains is less than in the  surrounding. When the spot moves upwards, the individual grains move downwards. }
      \label{figspot}
      \end{figure}

It is well known that a relatively general theory for  flow of granular media is still beyond reach. In part, this is due to  the richness of granular flow phenomena \cite{nedderman,gdr,beverloo,jenike} --- ranging from avalanche type behavior down inclined planes or granular surfaces to hopper discharges, granular gases, sheared flows in Couette cells or glassy type rheology. Other impediments to the development of a general framework to describe granular flow are the competition between static regions and flowing regions, and  the fact that flow zones are typically not much wider than a few particle diameters, so that continuum theories are questionable. Even though a general theory is still lacking, some more limited approaches aiming at describing the phenomena in a particular limit or in some special case have been relatively successful \cite{aranson01,boquet02,pouliquen06}. One particular recent proposal that appears to be quite successful for describing hopper flow and similar quasi-static flow problems is the so-called spot model introduced by Bazant and co-workers \cite{bazant06a,bazant06b,bazant07a,bazant07b}. The central idea of this approach, illustrated in Fig.~\ref{figspot}, is that the slow flow of dense random packings can be understood in terms of the drift and diffusion of relatively lower density regions of about 3-5 grains in diameter, called spots. While this idea is not unlike earlier approaches incorporating cooperative rearrangement effects, including the Soft Glassy Rheology phenomenology \cite{sollich97,sollich98} and the Shear Transformation Zone Theory \cite{falk98,lemaitre02,lemaitre04,bouchbinder07}, an attractive feature of the spot model is that it is more specific, permitting calculation of flow properties within a number of different experimental geometries and hence comparison to experiments or simulations. A second appealing aspect of the spot model is that it aims to merge the behavior at the scale of a few grains to the more ``engineering'' continuum-type approaches relating slow plastic flows to the stress fields (e.g. Mohr-Coulomb theory).

In this paper we will explore to what extent the spot model can be applied to the wide shear zones found in the so-called split-bottom geometries, i.e., cylindrical Couette cells in which the bottom consists of two rings, the inner of which rotates with the inner cylinder, the outer with the outer cylinder \cite{vanhecke03,vanhecke04,vanhecke06,chicago}. The shear zones found in such cells are exceptionally wide, much wider than the shear zones of order 5-10 particle diameters found in many other flow geometries \cite{gdr}. In addition, the  shear zone bends inwards towards the inner cylinder as a function of height in a way which appears to be determined largely by the balance of torques \cite{Unger}. 

A feature of the experimental data that made us initially optimistic that the spot model in its simplest form could be applied to these wide shear zones is an empirical finding, namely that the shear rate as a function of radius follows an error function profile with an effective width that depends on height \cite{vanhecke03}. As the spot model leads to a diffusion-type equation for the density of spots, fundamental solutions of the diffusion equation like Gaussians or error functions can be expected to emerge quite naturally for the spot density and the associated flow field. This intuitive observation motivated us to investigate the applicability of the spot model to these wide shear zone flows.
 
Contrary to our expectations, our main finding is that, without significant modifications or extensions, it is hard to reconcile the spot model with the main experimental observations on the wide shear zones. To guide our discussion with a concrete example, we will first briefly discuss in section 3 the results of a simple-minded layer approximation, which permits straightforward extension of the main features of the spot model in two dimensions to the three dimensional split-bottom geometry. This crude approximation always predicts quite narrow shear zones which essentially go straight up, rather than bend inwards as is found in experiments and according to the torque balance argument \cite{Unger}. As we shall see, these shortcomings are, however, not due to the inadequacy of the layer approximation underlying the implementation, but instead are intimately related to the basic structure of the drift-diffusion equation for spots. Indeed, in sections \ref{sectionflowrule} and \ref{sfssection} we explore in generality  which features of the spot model may be responsible for the incompatibility of the model with the wide shear zones.

There are essentially two different conceptual ingredients of the spot model which can be investigated separately: the more fundamental postulate that the flow can be analyzed in terms of a diffusion-type equation for the spot density, and the additional postulate that the drift vector in this diffusion equation can be calculated approximately from a Mohr-Coulomb type theory that predict stresses in materials at incipient yield. As we shall discuss, without additional modifications to the model, both features appear to be problematic for these wide shear flows. On the one hand, the drift-diffusion equation can be shown to be incompatible with the main experimental features of the wide shear zones. At the same time the co-axiality of the principal axes of the quasistatic granular stress and strain rate tensors, which according to theoretical arguments \cite{depken06} and recent simulations \cite{depken07} holds quite well in these wide quasi-static shear zones, is violated by Mohr-Coulomb stresses in the two dimensional Couette setup where the principal directions of the strain rate tensor are fixed by symmetry considerations.
In the outlook we will briefly mention some of the possible modifications and extensions that may be required to describe wide shear zones within a picture of drift and diffusion of spots.
  
\section{Essentials of the spot model}
As stated above, the essential idea of the spot model is that flow in granular matter is mediated by the opposing movement of  spots,  regions of excess free volume, on the order of 3-5 grain diameters in size. The excess free volume associated with a single spot is less than the volume of one grain. When a spot moves in one direction, a net particle flow is caused in the opposite direction. The movement of the spots is postulated to be the result of a combination of drift and diffusion; this is called the ``stochastic flow rule". Together with the incompressibility of the flow this leads to the following central equation
      \begin{equation} \label{eqnflowfield}      
        \vec{u}= \frac{L}{\Delta t} \left( - \dhat \rho+ \frac{L}{2} \vec{\nabla}\rho \right) ,     
      \end{equation}
relating the granular velocity field  $\vec{u}$ to the dimensionless spot density  $\rho$. Here $L$ is the spot size, $\Delta t$ the time it takes a spot to move a distance $L$, and $\dhat$ the normalized drift direction vector. The ``bare velocity'' $\vec{u}$ is smeared out by spatial convolution with a ``spot influence function'' to give the final flow field. Nevertheless, the essential features of the flow are captured by $\vec{u}$ and the drift vector $\hat{d}$, on which we therefore focus. Note that the time scale $\Delta t$ sets the velocity scale, so that only one parameter, the spot size $L$, governs the balance between the drift and diffusion terms. As we shall see, this simple feature is intimately tied to the fact that the spot model tends to give rise to narrow shear bands of width $L \approx$  3-5 grain diameters.

      The above expression can be thought of as the constitutive equation of the spot model: it postulates that granular flow results from the combined effect of a systematic drive (the drift term) and a random diffusion. The equation also illustrates that there are two sides to the spot model. On the one hand, Eq.~(\ref{eqnflowfield}) captures the essential idea that quasistatic granular flow can be captured completely in terms of the drift and diffusion of spots of reduced local density.  This idea can already be compared with known flow profiles if we know the drift direction vector $\dhat$ from simple symmetry arguments. E.g., in a two-dimensional hopper discharge it is argued \cite{bazant06a,bazant06b} that the drift direction vector $\dhat$ points straight up, as a result of gravity. In this case, the equation for the spot density reduces to a simple diffusion equation, with the vertical direction playing a role analogous to time. We will use similar symmetry arguments below in section \ref{sectionflowrule} to investigate the compatibility of the spot model with wide shear zones.
      
\begin{figure}
        \begin{center}
          \includegraphics[scale=0.45]{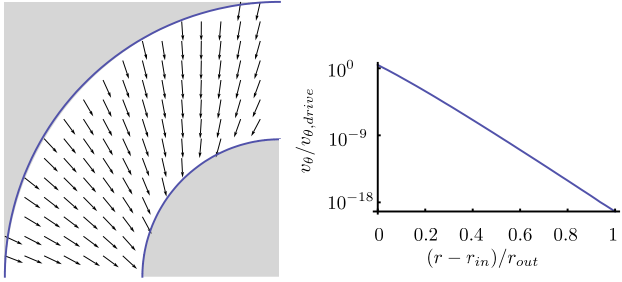}      
        \end{center}
      \caption{(left) Vector field for the drift direction vector $\dhat$ in one quarter of a cylindrical Couette cell cross section (inner and outer radii $r_{in}$ and $r_{out}$, respectively) driven with azimuthal velocity $v_{\theta, drive}$ at the inner boundary. For a Couette geometry in the absence of gravity, the drift direction vector is identical in every cross section. Note that while the drift has a large inwards radial component, this will be compensated by an equal and opposite diffusion, resulting in only azimuthal flow in the steady state. (right) The azimuthal flow velocity as a function of radius on a logarithmic scale, illustrating the almost purely exponential fall-off of the velocity, consistent with for example Ref.~\cite{gdr}.}
      \label{figcouetteresults}
      \end{figure}

The second side of the model is evident in nontrivial geometries, in which it is necessary to calculate the drift direction vector $\dhat$ explicitly. Calculation of $\dhat$ requires a theory of stresses in the material; it is here that the spot model in its simple form makes contact with traditional continuum methods from engineering. In order to deal with general 2d or quasi-2d cases, Kamrin and Bazant~\cite{bazant07a,bazant07b} have proposed that $\dhat$ can be determined from the Mohr-Coulomb plasticity theory (MCP theory). MCP theory posits that, immediately prior to plastic failure, granular materials are at incipient yield everywhere, i.e.~the material is just about to fail, collapse or exhibit plastic flow along at least two lines through each point of the material, called slip lines. The direction of these slip lines is determined by the stress fields and the value of the Coulomb friction coefficient $\mu$. Kamrin and Bazant assume that the stress tensor obtained from MCP remains valid in dense, quasistatic flows. They assert that spots perform a biased random walk along the slip lines, the bias being provided by the net force on a locally fluidized material region. This then determines the effective drift vector $\dhat$ from the stresses in the granular medium. Note that this formalism is essentially two-dimensional, a property that it inherits from the MCP theory. As an example of the type of result that can be achieved with the current, two-dimensional, spot-model we present both $\dhat$ and the azimuthal velocity in a two-dimensional Taylor-Couette disk in Fig.~\ref{figcouetteresults}; both results can be found in Ref.~\cite{bazant07a}.
      
\section{Example: crude layer approximation for the split-bottom geometry}
\begin{figure}
        \begin{center}   
	\includegraphics[scale=0.45]{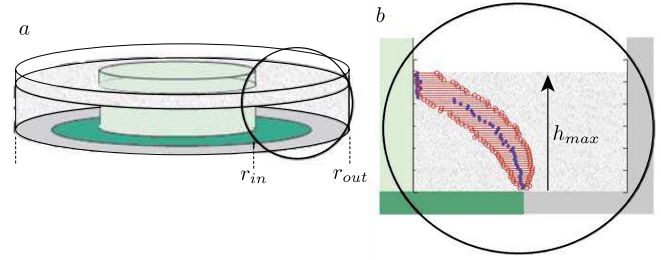}
        \end{center}
      \caption{(a) A schematic drawing of the split-bottom Couette geometry or Leiden geometry. The central part of the bottom rotates; the walls and the outer part of the bottom are stationary. (b) A cross-section of the flow. The center and width of the shear zone are indicated. Note that it is relatively wide and moves inwards and widens with increasing height. Figure taken from Ref.~\cite{vanhecke03} }
      \label{figsetup2}
\end{figure}

In order to give an idea of the difficulties of reproducing wide shear zones within the spot model, we briefly sketch the main result of a crude and minimal extension to three dimensions that we have developed to study the steady state shear profiles in the split-bottom Couette geometry of Fig.~\ref{figsetup2}  \cite{vanhecke03,vanhecke04,vanhecke06,chicago}. The shear bands observed in this system can exceed 50 grain diameters, much wider than what one typically finds. The location of the center of the shear band in this system is relatively accurately determined by a simple torque minimization argument applied to an infinitely thin shear zone \cite{Unger}. One would like the three-dimensional spot model to reproduce this and at the same time smear out these discontinuities to the relatively wide shear zones in the velocity field that  are found experimentally to be well fitted by an  error function profile.
            
There are a number of ways to expand the two-di\-men\-si\-onal spot model to a three-dimensional one.   One option is to replace the essentially 2d MCP theory with a 3d plasticity theory. Although possible, 3d plasticity theories come at the expense of a cumbersome mathematical framework. Below we will identify difficulties with the spot model inherent to both the stochastic flow rule and MCP. Though some plasticity theories may avoid the difficulties we identify with MCP, none can correct the issues associated with applying the stochastic flow rule to wide shear zones. Thus, for simplicity, we illustrate a generalization to 3d that builds on the 2d spot model incorporating MCP. We stress once more that we choose this simple and crude implementation here only for illustrative purposes --- the model can certainly be made more realistic but the arguments of sections 4 and 5 show that this will not change the basic tendency of the model to give straight narrow shear zones, as long as one sticks to Eq.~(\ref{eqnflowfield}). 
        
For the Couette geometry with symmetry along the axial direction, Kamrin and Bazant \cite{bazant07a} have already applied MCP. The corresponding $\dhat$-vectors are indicated in Fig.~\ref{figcouetteresults}. We build on these results by using a weak gravity approximation in which we consider the granular material to be composed of thin stacked slices. In each of these slices we assume that the two-dimensional picture holds, i.e.~we assume plane-strain boundary conditions. This means we can solve the two-dimensional spot model, based on MCP theory, in each slice to find the two-dimensional drift vector that would be there if this slice was the entire system \cite{footnote}. To this two-dimensional drift vector we then add a small vertical component due to gravity, which is constant both within each slice and between slices. The drift vector is thus
      \begin{equation}
        \hat{\vec{d} }=\dhat_\parallel (r,\theta,z) + \dhat_\perp \hat{z}.
      \end{equation}
Each slice is divided into two rings: an inner one that tends to rotate with the inner cylinder, and an outer one that tends to remain stationary in the lab frame. The boundary between the two rings is pinned at the split bottom but free to move with increasing height. Its position is determined self-consistently as we iterate upwards slice-by-slice.
      
\begin{figure}
        \begin{center}
          \includegraphics[scale=0.5]{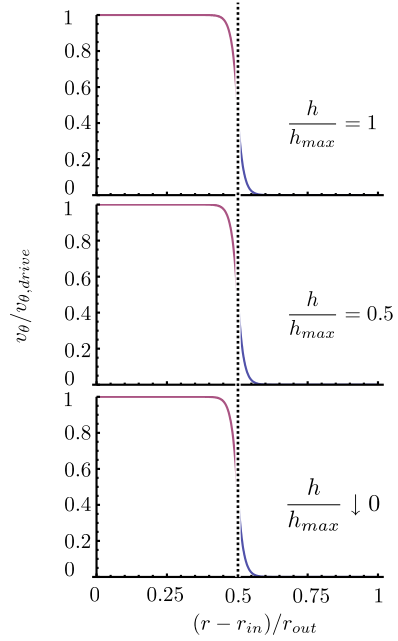}      
        \end{center}
      \caption{Results obtained in a three dimensional weak gravity approximation for the flow in a split-bottom Couette geometry. Plotted are the azimuthal velocities as a function of the rescaled radius for three different heights. The shear band remains narrow and centered above the split in the bottom of the cell. }
      \label{figresults}
      \end{figure}

Using this approach, one is  able to calculate the azimuthal velocity in a split-bottom Taylor-Couette setup as a function of both height and radius. As can be seen from  Fig.~\ref{figresults}, these results do not match the experimental observations of Fig.~\ref{figsetup2}: the shear band does not move inwards for larger heights and also shows no widening --- its width remains of order $L$, i.e. of the order of a few particle diameters. In the weak gravity limit, the fact that the shear bands remain of order $L$ can actually easily be understood from the result of Fig.~\ref{figcouetteresults}a: we already argued that as long as the $\dhat$-vector points mostly in the radial direction, the spot model constitutive equation (\ref{eqnflowfield}) predicts that the shear band width will be of order $L$ in the radial direction. We have explored various parameter regimes and other approximations, including a large gravity limit, but they always lead to essentially the same result:  we invariably find relatively narrow shear bands that shoot straight up. Since it was shown by Unger {\em et al.}~\cite{Unger} that the position of the shear band as a function of height can be obtained from a torque minimization argument, we consider it likely that it {\em is} possible to get the position of the shear band from a more accurate continuum stress calculation than we have done here. However, we focus here on the width of the shear band and now proceed to show that this is a generic feature of using spot model expression (\ref{eqnflowfield}) also in three dimensions, independent of which theory of stresses, MCP or otherwise, is used.
      
\section{Stress independent analysis}\label{sectionflowrule}

\begin{figure}
        \begin{center}
          \includegraphics[scale=0.65]{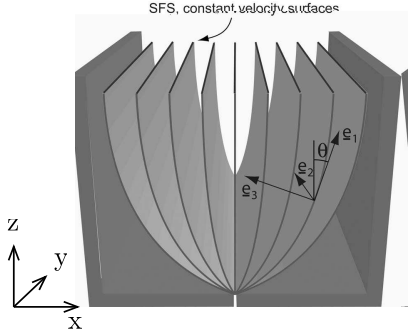}      
        \end{center}
      \caption{A number of shear free sheets in a linear shear cell, the shear free sheet basis is indicated with $e_{1,2,3}$. Figure taken from Ref.~\cite{depken07}.}
      \label{figSFS}
      \end{figure}

For simplicity, let us consider a linear split-bottom shear cell. As sketched in Fig.~\ref{figSFS}, this is an infinitely long rectangular container  filled with grains. The bottom of the container is split lengthwise, so that the two halves can be moved relative to each other. Simulations \cite{depken07} have confirmed that in this rectangular cell one obtains wide shear zones just as in the cylindrical Couette cell, but the geometry is somewhat easier to analyze due to the higher symmetry. If we denote the long dimension of the setup by $y$, the vertical dimension by $z$ and the cross-channel horizontal dimension by $x$, symmetry dictates the flow must be in the $y$-direction and that physical quantities like the velocity or spot density cannot  have any $y$-dependence. 

In the central vertical $y$-$z$-plane along which the two halves of the setup are sheared, both $\dhat_x$ and $\dhat_y$ must be zero due to symmetry, and since $\dhat$ is normalized $\dhat_z=1$ there. Neither in the experiments nor in the simulations is there any sign that there is a nonzero velocity $u_z$ at this center line; if we therefore impose $u_z=0$ at this center plane, we obtain from Eq.~(\ref{eqnflowfield}):
      \begin{equation}
        -\rho+\frac{L}{2} \frac{\partial \rho}{\partial z} = 0,     
      \end{equation}
which trivially leads to the exponential height dependence 
      \begin{equation}   
        \rho = A e^{\frac{2}{L} z},    \label{exponential} 
      \end{equation}
with $A$ an arbitrary constant. This equation illustrates clearly the tendency of the naive extension of the spot model to 3d to lead to exponential profiles of width of order $L$. For the split-bottom geometry, such an exponential dependence can obviously not be correct: in the experiments \cite{vanhecke03,vanhecke04,vanhecke06,chicago} as well as the simulations \cite{depken07},  wide shear zones at filling heights of $h_{max}=50$ grain diameters are studied. Even with $L=5$ Eq.~(\ref{exponential}) leads to an overall height variation of the spot density by an unrealistically large factor of order $e^{20}\approx 5\cdot 10^8$ --- to put  this in perspective, note that one spot is thought to contain a ``free'' volume of about a fifth of that of  a grain \cite{bazant06b}. Moreover, since the shear velocity in the $y$ direction away from the center line  is proportional to $\rho$, as there are no gradients in the $y$ direction, such an exponential dependence would imply an exponential variation of the $u_y$ velocity with height $z$, unless $\dhat_y$ is small and has a counterbalancing exponential height variation. Similar arguments apply if we analyze the extension of the profiles in the lateral (cross-channel) direction: for the profiles to be wide in the cross-channel direction, we can assume $\dhat_x \simeq - \alpha x$ with $\alpha \ll 1$ (this {\em Ansatz}  would lead to Gaussian variation of $\rho $ in the lateral direction) but the height dependence would remain exponential and close to that of (\ref{exponential}). Hence we conclude that independent of the precise ``flow rule" that relates $\dhat$ to the local stresses in the granular medium, the extension of the spot model, based on treating all components of $\dhat$ on an equal footing (as in Eq.~\ref{eqnflowfield}), is incompatible with the existence of wide shear zones. The reason for this is that the balance of the drift and diffusion term in Eq.~(\ref{eqnflowfield}) is governed by the single  spot diameter length scale $L$ --- given that the spot size is hypothesized to be of order $3-5$ grain diameters this almost inescapably leads to narrow shear bands  of  only up to ten grain diameters wide. 
                    
\section{Co-axiality and Shear Free Sheets}\label{sfssection}

The above discussion is independent of which particular flow rule is adopted for the connection between the stresses and the drift vector $\dhat$. It is nevertheless of interest to go back and discuss the relation between the principal stress and shear directions on the basis of what we know from recent theory and simulations \cite{depken06,depken07,bazantprivate}, and to explore the implications for theories on granular flow.

In the original formulation of the spot model, the so-called co-axiality flow rule is explicitly rejected and replaced by a flow rule that builds on MCP theoy. Co-axiality means that the principal axes of the stress tensor and the strain rate tensor are co-axial, i.e., aligned. However, according to the the analysis of wide shear flows of Depken {\em et al.} \cite{depken06}, there are various reasons to believe that co-axiality is actually a crucial feature of these wide shear zones; later simulations confirmed this picture surprisingly well \cite{depken07}. 
      
In order to explore what this implies, we need to explain briefly the Shear Free Sheet (SFS) basis introduced by Depken and co-workers. A SFS is a surface of constant velocity; for the case of the rectangular split-bottom geometry the SFS's are sketched in Fig.~\ref{figSFS}. These sheets naturally define the orthogonal basis spanned by the unit vectors $\hat{e}_1$ and $\hat{e}_2$ in the SFS (with $\hat{e}_2$ is taken in the direction of flow)  and $\hat{e}_3$ perpendicular to the sheet. Since by construction the velocity is constant within each sheet, the strain rate within each SFS is also zero. An easy calculation then shows us that two of the principal directions of strain are at angles of $\pi/4$ relative to $\hat{e}_2$ and $\hat{e_3}$, while the third principal direction is the same as $\hat{e}_1$. In line with the theoretical arguments \cite{depken06}, recent simulations \cite{depken07,bazantprivate} confirm that co-axiality holds in these wide shear zones, and hence that the orientation of the axes of principal stress are fixed by the SFS's.

      \begin{figure}
      \begin{center}
        \includegraphics[scale=0.7]{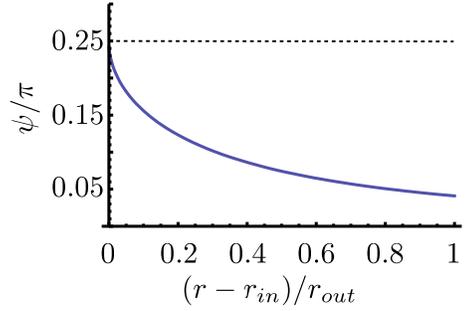}     
      \end{center}
        \caption{(solid curve) The value of $\psi$, the angle between the radial axis and the axis of minor principal stress, as a function of radius in a planar Couette setup. According to the results of Ref.~\cite{depken06} this should always be close to $\frac{1}{4}\pi$ (dashed line) or $\frac{3}{4} \pi$ in a wide shear zone.}
        \label{figpsi}
      \end{figure}

To illustrate how different the Mohr-Coulomb picture is from co-axiality in wide shear zones, let us return briefly to the approximation in which we think of the cylindrical Couette cell as being built up from slices of the  two-dimensional (disk-like) Couette setup. Since the flow is azimuthal, the SFS's in each slice are just concentric circles. Since the principal directions of strain rate are at $\pi/4$ angles to these sheets, and since the shear zones bend only slightly inwards towards the inner radius with increasing height, they make an angle close to $\pm \pi/4$ with the azimuthal direction. Due to co-axiality we know that this must also be true for the principal stress directions. This means that the angle between the radial axis and the minor principal stress axis is in reality always close to $\frac{1}{4}\pi$ or $\frac{3}{4}\pi$ depending on which of the two is the major and which is the minor principal stress direction. In the two-dimensional Mohr-Coulomb theory, however, the principal stress axes point in very different directions throughout most of the layer, as Fig.~\ref{figpsi} illustrates. In other words, any plasticity theory that presumes a violation of co-axiality, does not appear to be a viable starting point for the description of wide shear zones. Any theory based on drift and diffusion of spots will have to build in, or self-consistently lead to, co-axiality in the wide shear zones.
      
\section{Outlook}      
In putting our results into perspective we would like to stress that the spot model was not formulated specifically to apply to wide shear zones. In spite of this, the error function shear profiles found experimentally and the success of the model in capturing Gaussian hopper flow profiles led us to become optimistic that the model could capture the wide shear zones as well.  Nevertheless, our ana\-ly\-sis shows that present formulations of the spot model cannot capture the physics of the wide shear zones observed in split-bottom experiments and simulations. The main reason is that the model in its current form is based on the interplay and balance of diffusion and drift. As Eq.~(\ref{eqnflowfield}) clearly shows, this balance is governed by the single length scale $L$ of order 3-5 grain diameters. As a result, in its present form the structure of the model is such that it leads to localization of the shear bands on this same scale $L$.

More generally, our analysis brings up the question whether static stress considerations can be used to determine the properties of wide shear zones: In Mohr-Coulomb theory and in the present spot model, one attempts to calculate the flow fields from static stress fields in conjunction with an incipient yield postulate, but the co-axiality of the stress and strain rate tensor suggests instead a picture in which the flow ``self-organizes'' to a co-axial state.

To be more specific: we have based our discussion on the central result, Eq.~(\ref{eqnflowfield}), for the connection between the drift vector $\dhat$ and the velocity field. In their discussion of the applicability of the spot model to other flow geometries, Kamrin and Bazant \cite{bazant07a} have independently advanced the idea that spot drift must align with special surfaces (related to ``slip line admissability'') \cite{kensremarks}. If indeed we postulate that the spot drift is orthogonal to the $z$-direction in the central $yz$ plane of Fig.~\ref{figSFS}, then clearly the main problem --- an unrealistic exponential variation of the spot density $\rho$ with height --- dissolves. If $\dhat$ is not along the $z$-direction in the central $yz$ plane, then by symmetry it must be zero. Kamrin and Bazant in fact already interpreted a region with a null drift
vector to be an indication that the spot mechanism is weak (a similar situation occurs in plane shear and inclined plane flow \cite{bazant07a}). So from that perspective too, one unfortunately has to conclude that wide shear zones include physics that go beyond the spot approach. Other possible routes to extending the spot idea to accommodate wide shear zones might be to introduce spot generation and annihilation terms \cite{martin} (i.e.~regions of high plastic strain may act like a source of spots); to allow for evolution in spot size, thereby implicitly introducing another length scale; or to use an approach as suggested in Ref.~\cite{bazant07a}  that reconciles Bagnold type rheology with the physics of the stochastic flow rule.

\section*{Acknowledgments}
We are grateful to Martin Bazant and Ken Kamrin for openly sharing their thoughts on the strengths and weaknesses of the spot model with us and for extensive comments on an earlier version of this manuscript. We also thank Martin Depken and Martin van Hecke for discussions, encouragement and critique.

\end{document}